# Incorrectness of analytical solution of Schrodinger equation for diatomic molecule with shifted Tietz-Wei potential

I.H. Umirzakov
Institute of Thermophysics, Lavrentev prospect, 1, 630090 Novosibirsk, Russia
e-mail: cluster125@gmail.com

**Abstract** It is shown that: the analytical solution of the stationary Schrodinger equation for diatomic molecules with the shifted Tietz-Wei potential, which was reported in the paper "Exact and Poisson summation thermodynamic properties for diatomic molecules with shifted Tietz potential, A N Ikot, W Azogor, U S Okorie, F E Bazuaye, M C Onjeaju, C A Onate and E O Chukwuocha, *Indian Journal of Physics* (2019)", is incorrect; the analytical expressions for the rotational-vibrational energy spectrum for diatomic molecule obtained from the incorrect analytical solution of the Schrodinger equation is incorrect; and free energy, internal energy, entropy and specific heat of cesium dimer $Cs_2$ at the electronic state $3^3\Sigma_g^+$ calculated from the incorrect expression for the rotational-vibrational energy spectrum are incorrect.

The shifted Tietz-Wei interaction potential for the diatomic molecules is given by [1,2]

$$V(r) = D_e \cdot \frac{2(c-1)e^{-b(r-r_e)} - (c^2-1)e^{-2b(r-r_e)}}{[1-ce^{-b(r-r_e)}]^2}, \qquad (1)$$

where $D_e$ is the dissociation energy, $D_e > 0$, $r$ is the interatomic separation, $r \geq 0$, $r_e$ is the equilibrium bond length, $r_e > 0$, $c$ is the optimization parameter, $b = \gamma(1-c)$, $b > 0$, $\gamma$ is the Morse constant and $\gamma > 0$.

1. The detailed analysis of derivation of Eq. 8 from Eq. 3 in [2] shows that:

the definition of the variable $y$ by $y = (1-ce^{-\alpha r})^{-1}$ is incorrect because, particularly, the variable of exponent $-\alpha r$ has the dimensionality of distance while it must be dimensionless, and the correct definition of $y$ is $y = (1-ce^{-b(r-r_e)})^{-1}$; and

Eqs. 9b and 9c in [2] are incorrect and it is necessary to replace these equations by the correct ones:

$$P = \frac{2\mu D_e}{\hbar^2 b^2 c^2}(c-1)^2 + \frac{J(J+1)D_2}{b^2 r_e^2 c^2}, \qquad (9b)$$

$$Q = \frac{2\mu D_e}{\hbar^2 b^2 c^2}(c^2-1) + \frac{J(J+1)}{b^2 r_e^2 c^2}(cD_1 - D_2). \qquad (9c)$$

2. The detailed analysis of derivation of Eq. 10 from Eq. 9 in [2] shows that the term $\sqrt{1/4 - P}$ in Eqs. 10 and 13-20 in [2] are incorrect, and it is necessary to replace this incorrect term by correct one, which equal to $\sqrt{1/4 + P}$. Besides the expression for the parameter $\sigma$ given after Eq. 21 [2] is incorrect, and this parameter in Eq. 21 [2] for the rotational-vibrational energy spectrum of diatomic molecule must be defined from the correct one which is given by $\sigma = 1/2 - \sqrt{1/4 + P}$. Hence, the analytical solution of the Schrodinger equation for diatomic molecules with the shifted Tietz-Wei potential and the analytical expression for the rotational-vibrational energy spectrum of diatomic molecule obtained in [2] are incorrect, and the free energy, internal energy, entropy and specific heat of cesium dimer $Cs_2$ at the electronic state $3^3\Sigma_g^+$ calculated from above incorrect expressions are incorrect.